\documentclass[aip,rsi,reprint,graphicx]{revtex4-1} 

\usepackage{amsmath, newtxtext, newtxmath}

\usepackage[pdftex]{graphicx}
\usepackage{dcolumn}
\usepackage{bm}
\usepackage{soul}

\usepackage{bm}
\usepackage{soul}

\usepackage{hyperref}
\hypersetup{
    colorlinks=true,        
    linkcolor=blue,          
    citecolor=blue,        
    filecolor=magenta,      
    urlcolor=blue           
}
\usepackage{color}
\newcommand{\textbl}{\textcolor{black}}

\draft 

\begin{document}


\title{High-speed 100 MHz strain monitor using fiber Bragg grating and optical filter for magnetostriction measurements under ultrahigh magnetic fields} 



\author{Akihiko Ikeda}
\email[e-mail: ]{ikeda@issp.u-tokyo.ac.jp}
\author{Toshihiro Nomura}
\author{Yasuhiro H. Matsuda}
\email[e-mail: ]{ymatsuda@issp.u-tokyo.ac.jp}
\affiliation{International MegaGauss Science Laboratory (IMGSL), Institute for Solid State Physics (ISSP), University of Tokyo, Kashiwa, Japan}
\author{Shuntaro Tani}
\author{Yohei Kobayashi}
\affiliation{Laser and Synchrotron Research Center (LASOR), Institute for Solid State Physics (ISSP), University of Tokyo, Kashiwa, Japan}
\author{Hiroshi Watanabe}
\affiliation{Graduate School of Frontier Biosciences, Osaka University, Suita, Japan}
\author{Keisuke Sato}
\affiliation{Department of Natural Science, Ibaraki National College of Technology, Hitachinaka, Japan}

\date{\today}

\begin{abstract}

A high-speed 100 MHz strain monitor using a fiber Bragg grating, an optical filter, and a mode-locked optical fiber laser has been devised, which has a resolution of $\Delta L/L\sim10^{-4}$.
The strain monitor is sufficiently fast and robust for the magnetostriction measurements of magnetic materials under ultrahigh magnetic fields generated with destructive pulse magnets, where the sweep rate is in the range of 10--100 T/$\mu$s.
As a working example, the magnetostriction of LaCoO$_{3}$ was measured at room temperature, 115 K, and 7$\sim$4.2 K up to a maximum magnetic field of 150 T. The smooth $B^{2}$ dependence and the first-order transition were observed at 115 K and 7$\sim$4.2 K, respectively, reflecting the field-induced spin state evolution.

\end{abstract}


\maketitle 

\section{Introduction}

Magnetostriction, changes in the lattice parameters of magnetic materials in response to an external magnetic field, is an indispensable phenomenon in the research on magnetic materials, where strong spin-lattice coupling plays a crucial role.

Strain monitoring using fiber Bragg grating (FBG) is now widely employed in non-destructive testing applications.\cite{Kersey}
FBG is an optical fiber with a Bragg grating placed at the fiber core.
The Bragg grating is formed by the modulation of the refractive index of the fiber core in the axial direction.
Light with a Bragg wavelength $\lambda_{\rm{B}}=2nd$ is reflected back by the Bragg grating, where $n$ and $d$ are refractive index and period of the Bragg grating, respectively.
$\lambda_{\rm{B}}$ shifts in proportion to the change in the length $\Delta L$ of the FBG.
Thus, the $\Delta L/L$ of materials coupled to the FBG can be measured by monitoring $\Delta \lambda_{\rm{B}}/\lambda_{\rm{B}}$.
FBG is presently an established technique used for telecommunication and non-destructive testing. \cite{Othonos}
Hence, FBG is available at low cost and demonstrates high stability.

FBG-based high-resolution magnetostriction measurements have been carried out on various magnetic materials up to 100.75 T generated using non-destructive pulse magnets with pulse durations of the order of milliseconds.\cite{Jaime_PNAS2012, Moaz, Rotter}
The reported system works at 47 kHz with a resolution of $\Delta L/L\sim10^{-7}$.\cite{Daou}
Magnetostriction measurement using the FBG technique is advantageous in several respects in comparison with other techniques such as a capacitance bridge or strain gauge.\cite{Daou}
\textbl{FBG} is immune to electromagnetic and mechanical noise and does not require calibration, which limit the resolution of $\Delta L/L$ in the aforementioned methods.\cite{Doerr}

At ultrahigh magnetic fields beyond 100 T, remarkable magnetic transitions have been found in the magnetization measurements. Examples are the end of the half-magnetization plateau of an orthogonal dimer spin system, \cite{SCBO} the phase transition from the $\alpha$-phase to $\theta$-phase in solid oxygen,\cite{NomuraPRL2014} the spin crossover between the low-spin phase, high-spin phase, and possible excitonic insulator phase in perovskite cobaltites, \cite{Ikeda201601, Ikeda201602} and the transition from a Kondo insulator  to the heavy fermion state.\cite{TerashimaJPSJ}
All these are expected to be accompanied by significant magnetostriction owing to the strong spin-lattice coupling.
Hence, magnetostriction measurements under ultrahigh magnetic fields are desirable to gain comprehensive insight into these phenomena.

\begin{figure*}
\includegraphics[scale=0.45]{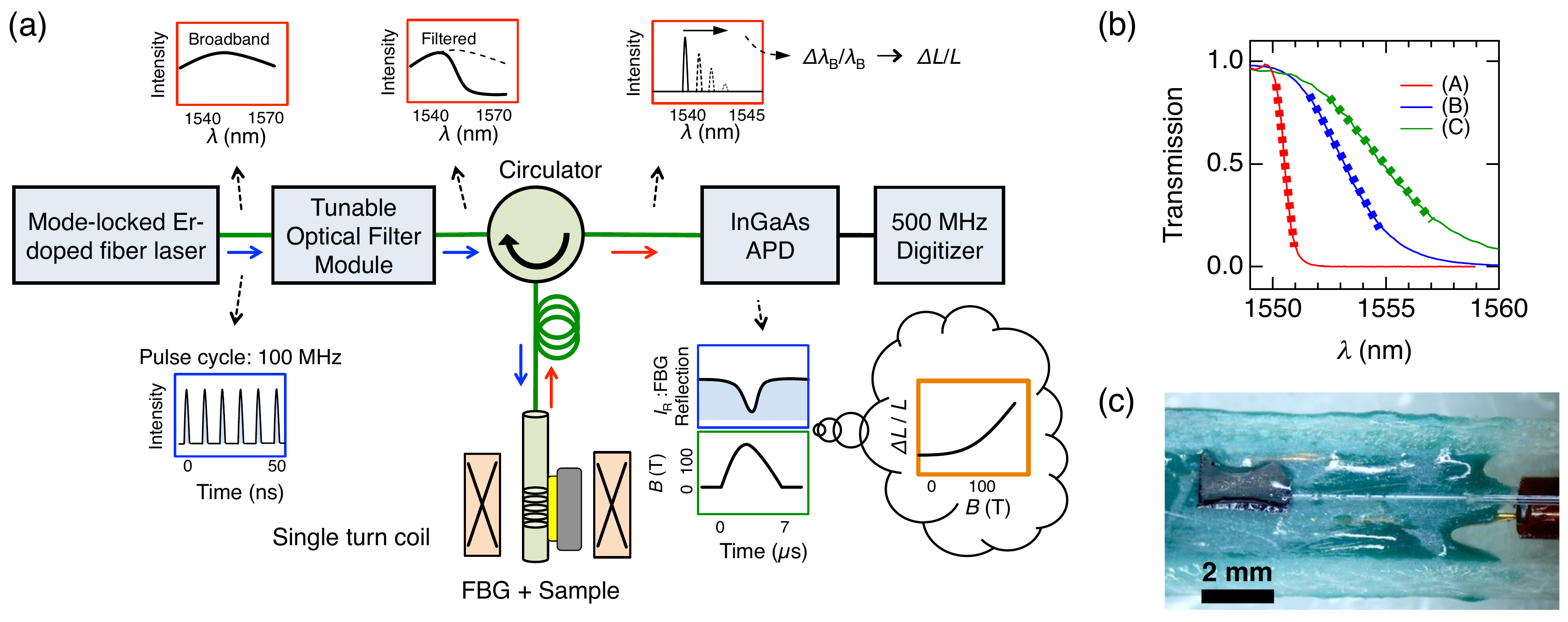}
\caption{\label{schem}(a) A schematic of the high-speed strain monitoring system. Green lines indicate the single-mode optical fiber line. The red- and blue-lined rectangles indicate the spectral and temporal profile of the propagating light at the respective positions. (b) Transmission spectra of the optical filters (A) to (C). Refer to the main text for the details of the filters. (c) Photograph of an FBG fiber glued to a single-crystalline LaCoO$_{3}$. }
\end{figure*}

Generating magnetic fields beyond 100 T necessarily requires techniques employing destructive pulse magnets, such as single turn coil (STC) and magnetic flux compression,\cite{Herlach} where the field sweeps typically at 10--100 T/$\mu$s accompanied by a large electromagnetic noise.
Thus, physical measurements under ultrahigh magnetic fields require properties such as (1) single-shot measurement (2) fast measurement at $f> 10$ MHz and (3) resistance to electromagnetic noise.
Examples of conventionally established measurement techniques are the magneto-optical spectroscopy, Faraday rotation measurements, \cite{MiyataPRL} inductive magnetization measurement using pickup coils, \cite{TakeyamaM} and radio frequency (RF)-based conductivity measurements. \cite{SuyeonJPSJ}
However, in electrical measurements, the electromagnetic noise is always a serious problem.

An FBG-based magnetostriction measurement at 100 MHz with a strain resolution of $\Delta L/L<10^{-4}$ was reported recently.\cite{Rodriguez}
In the system, a technique that converts the spectrum to the time domain is utilized, where the spectrum of the FBG reflection is recorded at 100 MHz.
The conversion technique relies on a long dispersive optical fiber with a length of \textbl{30--100 km} and a high-speed digitizer with a bandwidth of 25 GHz.
The system is significantly advantageous in the unambiguous determination of the spectrum of an FBG reflection.
On the other hand, the system requires high quality optical and electrical components.
Further, to improve its resolution, space and simplicity of operation need to be sacrificed.

In this paper, we present an unconventional system of a high-speed 100 MHz FBG strain monitor. 
\textbl{The shift of the FBG reflection peak was converted to the intensity of the reflected light by the FBG by using the cut-off region of the optical filters.\cite{Tsuda}
A 100-MHz mode-locked fiber laser, a broadband and very bright light source, was employed.
By the combination of those two methods, we succeeded in measuring the strain of solids at 2000 times greater repetition rate than that of the previous report.\cite{Daou}
The repetition rate of the measurement is fast enough to be applied to the magnetostriction measurement at $B>100$ T with the destructive pulse magnet whose pulse duration is limited to several microseconds.}
The system complemented the conventional system \cite{Rodriguez} with its compactness and simplicity of operation \textbl{due to the alternative methodology}.
To demonstrate the capability of the system, we carried out the magnetostriction measurements of a single crystalline LaCoO$_{3}$ at room temperature, 115 K, and $7\sim4.2$ K up to a maximum magnetic field of 150 T.
\textbl{This is the first report on successful measurements of the magnetostriction at above 100 T.}

\section{Method}

\textbl{The principle of the measurement of the magnetostiction $\Delta L/L$ of solids as a function of the magnetic field $B$ is as follows. The time evolution of the longitudinal magnetostriction $\Delta L/L(t)$ of the sample is first converted to the $\Delta \lambda_{\rm{B}}/\lambda_{\rm{B}}(t)$ of the FBG. Then, using an optical filter, $\Delta \lambda_{\rm{B}}/\lambda_{\rm{B}}(t)$ is converted to the change of the light intensity $\Delta I_{\rm{R}}(t)$. $\Delta I_{\rm{R}}(t)$ is measured along with the pulsed magnetic field $B(t)$. With proper inverse transformation, one can obtain $\Delta L/L(B)$. To keep the transformation simple, in the present study, the measurements are performed in the region where the following relation holds, $\Delta L/L(t) \propto \Delta \lambda_{\rm{B}}/\lambda_{\rm{B}}(t) \propto \Delta I_{\rm{R}}(t)$. Further details are the followings.}

\begin{figure*}
\includegraphics[scale=0.6]{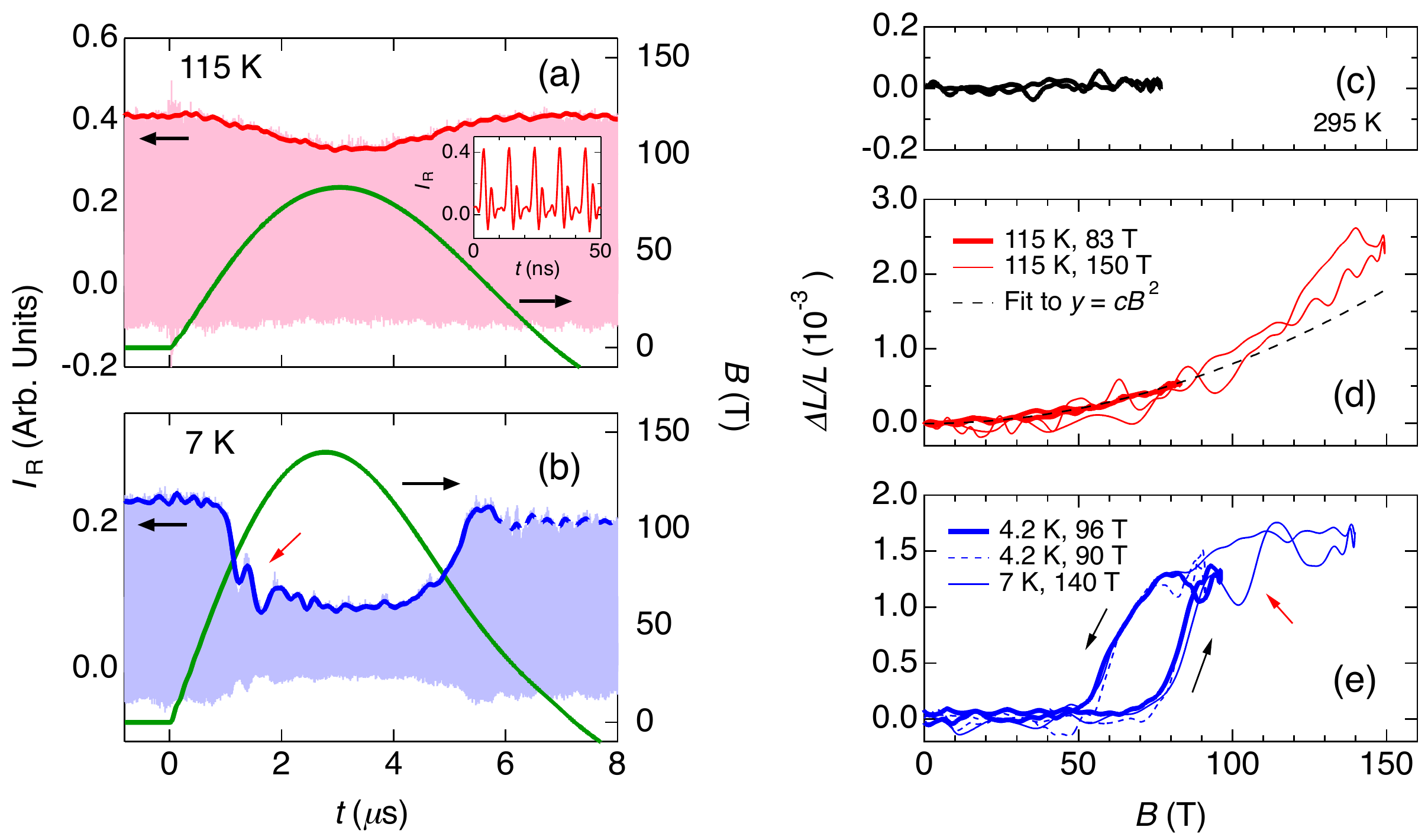}
\caption{\label{result}Time evolution of the pulsed magnetic fields and the intensity of the FBG reflection $I_{\rm{R}}$ from the FBG fiber glued to LaCoO$_{3}$ at (a) 115 K and (b) 7 K.
The inset shows the temporal magnification of the intensity of the FBG reflection.
The obtained longitudinal magnetostriction of LaCoO$_{3}$ along the pseudo-cubic [001] axis up to (c) 80 T at 295 K , (d) 85 T and 150 T at 115 K, and (e) 90 and 95 T at 4.2 K and up to 140 T at 7 K.
The arrows pointing up and down correspond to the upward and downward sweeps of the magnetic fields.}
\end{figure*}

The schematic diagram of the devised high-speed strain monitoring system is shown in Fig. \ref{schem}(a).
As a broadband light source, a mode-locked Er-doped fiber laser with a repetition rate of 100 MHz was employed.
It had a continuum spectrum centering at $\sim1550$ nm, with a width of about 60 nm and an output power of 750 $\mu$W.
\textbl{Note that it is unfavorable to use amplified stimulated emission (ASE) sources, namely superluminescent diodes, instead of the mode-locked fiber laser.
Whereas the spectrum of the ASE sources looks broadband with stability at a slow time scale, it is rather fluctuating at RF time scales, which insufficient for the current requirement of the stable broadband source at RF time scale.}
The light source was filtered with an optical filter. 
By changing the optical filter between (A) to (C) (Koshin Kogaku (A) 1560SPF, (B) 1551BPF and (C) KD-3091), the sensitivity and the dynamic range of $\Delta \lambda_{\rm{B}}$ could be varied, whose spectrums are shown in Fig. \ref{schem}(b).
The optical filter (B) was mainly used in the present study.
The cut-off wavelength of optical filters was adjustable for $\pm\sim10$ nm around $\sim$1555 nm with a tunable filter module (Koshin Kogaku TFM/FC). \textbl{One may need to retry the measurement with a different optical filter when the measurement is failed either due to the limited dynamic range or the poor sensitivity of the optical filter.}

The filtered light source illuminated the FBG after passing through an optical circulator.
The FBG had a length of $\sim2.5$ mm.
The width of the Bragg peak was $<0.7$ nm.
The light reflected by the FBG was guided by the optical circulator and detected with an InGaAs avalanche photodetector (APD) (Thorlabs APD430C) with a conversion gain and bandwidth of $1.8\times10^{5}$ V/W and 400 MHz, respectively, and an oscilloscope (Lecroy HDO4054) with a bandwidth of 500 MHz. 
The typical intensity of the light reflected by the FBG was $< 3$ $\mu$W.

A bare glass fiber with the FBG was glued to the single-crystalline perovskite LaCoO$_{3}$ along the pseudo-cubic [001] axis with epoxy (Stycast 1266), as shown in Fig. \ref{schem}(c).
\textbl{LaCoO$_{3}$ is a widely investigated material for its unusual spin crossover from the low-spin (LS) ground state to unidentified excited spin state.\cite{Ikeda201601}}
The optical fiber had an SMF-28e coating, except for the region with the FBG.
$\Delta L/L$ of the sample was then converted to $\Delta \lambda_{\rm{B}}/\lambda_{\rm{B}}$, with the relation $\Delta \lambda_{\rm{B}}/\lambda_{\rm{B}}=c_{1}\Delta L/L
+c_{2}\Delta T$, where $c_{1}=1-(n^{2}/2)[P_{12}-\nu(P_{11}+P_{12})]\simeq 0.76$ and $c_{2}=\alpha(T)+(dn/dT)/n$. \cite{Kersey}
$P_{ii}$, $\nu$, $n$, and $\alpha(T)$ are components of the strain-optic tensor, Poisson ratio, refractive index of the optical fiber, and linear thermal expansion, respectively.
It is reasonable to assume $\Delta \lambda_{\rm{B}}/\lambda_{\rm{B}}=0.76\Delta L/L$ in the measurements at fixed temperatures.\cite{Daou}

The shift of $\lambda_{\rm{B}}$ was converted to the change in the intensity of the FBG reflection $\widetilde{\Delta I_{\rm{R}}}=-(I_{\rm{R, }\it{B}\rm{>0}}-I_{\rm{R, }\it{B}\rm{=0}})/I_{\rm{R, }\it{B}\rm{=0}}$ as a result of the filtered light source.
Here, $I_{\rm{R}}$ was assumed to be the amplitude of the signal at 100 MHz detected by the APD.
The wavelength region of the optical filter where the relation of $\Delta \lambda_{\rm{B}}/\lambda_{\rm{B}}=a \widetilde{\Delta I_{\rm{R}}}$ held, was used in the experiment, with $a$ being a constant dependent on the optical filters (A) to (C), shown as the slope of the dotted region in Fig. \ref{schem}(b).
\textbl{For the optical filter (B), $a=3.87\times10^{-3}$.}
$\Delta L/L$ of the sample as a function of the magnetic field was deduced using the obtained time evolution of $I_{\rm{R}}$ and the magnetic field, as shown in Fig. \ref{schem}(a).

The ultrahigh magnetic fields were generated using a horizontal and a vertical STC in the Institute for Solid State Physics, University of Tokyo.\cite{Miura} The cryostats used were He-flow \cite{Amaya} and He-bath types \cite{TakeyamaM} for the horizontal and the vertical STCs, respectively.

\section{Results and discussion}
As a working example, we measured the magnetostriction of perovskite LaCoO$_{3}$ at room temperature, 115 K, and $7\sim4.2$ K, up to a maximum magnetic field of 150 T.
The obtained magnetic field and $I_{\rm{R}}$, as functions of time, are shown in Figs. \ref{result}(a) and \ref{result}(b).

At 115 K, as shown in Fig. \ref{result}(a), $I_{\rm{R}}$ decreases gradually with an increasing magnetic field and recovers to the original intensity with a decreasing magnetic field.
On the other hand, at 7 K, as shown in Fig. \ref{result}(b), $I_{\rm{R}}$ showed a sudden decrease with an increasing magnetic field at well after $t=0$ $\mu$s, and suddenly recovered back to the initial intensity with a decreasing magnetic field.
Subsequently, the amplitude of $I_{\rm{R}}$ decreased once more at $\sim6$ $\mu$s, which could be attributed to the partial detachment of the FBG that does not occur with magnetic field pulses of  $B_{\rm{max}}<100$ T.

A series of calculated $\Delta L/L$ are shown in Figs. \ref{result}(c)-\ref{result}(e).
At room temperature, as shown in Fig. \ref{result}(c), $\Delta L/L$ seemed not dependent on the magnetic field \textbl{within the present experimental error}. 
In Fig. \ref{result}(d), $\Delta L/L$ at 115 K showed a $B^{2}$ dependence up to 100 T and deviated upward at above 100 T.
The $B^{2}$ dependence of $\Delta L/L$ up to 40 T was also reported previously.\cite{Sato2008}
In Fig. \ref{result}(e), $\Delta L/L$ showed good agreement with the previous reports on the first-order phase transition with hysteresis and the $\Delta L/L$ at the transition of $\sim1.5\times 10^{-3}$. \cite{Moaz, Rotter}
On the other hand, at 115 K, our data at 35 T quantitatively differed by a factor of $\sim1/3$ from the reported data of longitudinal $\Delta L/L$ along the [111] axis at 100 K. \cite{Sato2008}


\textbl{Here, we discuss the noise level of the present measurements in various circumstances. 
To be brief, the typical noise level on $\Delta L/L$ is of the order of $10^{-4}$, judging from the data in Fig. \ref{result}(c)-\ref{result}(e) with the following analysis. 
Following analysis of noise is typically applicable to the optical filter (B) and may differ depending on the optical filters. 
First, as can be judged from the data at room temperature in Fig. \ref{result}(c), we may conclude that the noise level of the preset set up is $\sim5\times10^{-5}$. 
At the room temperature, the sample shows no appreciable $\Delta L/L$.  \cite{SatoPhD}
Therefore, the noise of the data in Fig. \ref{result}(c) can be considered to be the white noise of the current system under the pulsed magnetic filed. 
Also we may state that there is no influence of the magnetic field to the FBG reflection up to 80 T due to such as magneto-optical effects, being in good agreement with previous studies.\cite{Daou, Rodriguez} }

\textbl{Here we discuss the source of the white noise in Fig. \ref{result}(c). 
One of the main source of the white noise is considered to be the intrinsic temporal fluctuation of the intensity of the laser light source, based on the fact that the yield of the white noise is identical in the temporal region before and after the pulsed magnetic field is started.
This being true, then, eliminating this fluctuation from the signal by employing the differential detection of $\Delta I_{\rm{R}}$ will significantly improve the signal-to-noise ratio.
}

\textbl{Another factor affecting the noise level in the measurement at low temperatures is the cooling scheme. 
The noise level showed a dependence on the cooling scheme, namely the He-bath type cryostat used for shots with $B_{\rm{max}}<100$ T (Thin solid and dashed curves in Fig. \ref{result}(d) and \ref{result}(e)), and the He-flow type cryostat used for $B_{\rm{max}}>100$ T (Thick solid curves in Fig. \ref{result}(d) and \ref{result}(e)). 
With the He-bath type cryostat, the noise level is as small as that of the room temperature data in Fig. \ref{result}(c). 
On the other hand, with the He-flow cryostat, the noise level become $\sim10^{-4}$ before and during the pulsed magnetic field is generated (Thick solid curves in Fig. \ref{result}(d) and \ref{result}(e)).
This fact may indicate that the FBG and the sample is mechanically unstable against the intense He flow within the cryostat, which may be eliminated by fixing the sample more tightly to the cryostat. 
}

\textbl{Lastly in the noise analysis, we mention the large bounce at $\sim3$ MHz indicated by the red arrows in Fig. \ref{result}(b) and \ref{result}(e), occurred just after the first-order phase transition.
This bounce may originate either in intrinsic or extrinsic effects, such as mechanical vibration of the sample or the photoelasticity of the FBG fiber under elastic loading, respectively.
The latter effect is safely excluded by checking the FBG spectrum during the cooling process of the sample from the room temperature to liquid N$_{2}$ or He temperature at zero magnetic field. 
During cooling the sample, the FBG peak shows about shift of several nanometers with no deformation of the FBG spectrum and the decrease of the reflection intensity.
Besides, strain-induced photoelasticity in the FBG fiber is not reported in previous literature.\cite{Daou, Rodriguez}
Therefore, the bounce is considered to be due to the elastic vibration of the sample at its resonant frequency along the FBG initiated by the first-order phase transition.
Assuming that the speed of sound in the material and the sample size are of the order of 5 km/s and 2 mm, the fundamental mode of the elastic vibration of the sample should have the frequency of $2.5$ MHz, being in good agreement with the observation of $\sim3$ MHz.
On this basis, we regard this bounce as the intrinsic length change of the sample and discard the possibility of the change of the noise level.
In conclusion for the noise analysis, we have the typical noise level of $\sim10^{-4}$ in the current measurements.
The noise level should depend on the situation of the measurements and still have much room to improve.}

\textbl{Here we discuss the physical origins of the measured lattice changes.
The above results shown in Fig. \ref{result}(c)-\ref{result}(e) can be understood in terms of the magnetic field induced spin crossover of Co$^{3+}$ in LaCoO$_{3}$ from the low-spin (LS) ground state to excited spin states at high magnetic fields.
In Fig. \ref{result}(d), the $B^{2}$ dependence of $\Delta L/L$ at 115 K may indicate the elastic lattice effect \cite{Sato2008} where the repulsive interactions between LS and the excited spin state is inclined to prevent the occupation of the excited spin states in the LS background.
Once the excited spin state is populated by a fraction of Co$^{3+}$ in the solid at low magnetic fields, they start to increase their occupation cooperatively at higher magnetic fields due to the effective attractive interactions between the excite spin states.
The deviation from the $B^{2}$ dependence at $B>100$ T may infer the acceleration of the cooperative spin crossover at even large occupation of the excited spin state.}

\textbl{In Fig. \ref{result}(e), the first order spin-state transition takes place from the LS to an unidentified excited spin state at high magnetic fields.
The observation is quantitatively in good agreement with previous  reports on magnetostriction measurements with the non-destructive pulse magnets below 100 T \cite{Moaz, Rotter} and the magnetization measurements at above 100 T.\cite{Ikeda201601}
It is not clear at this moment what is the mechanism of the field-induced spin crossover. 
As a plausible mechanism, several theoretical studies are arguing the field-induced Bose-Einstein condensation of excitons,\cite{Tatsuno, Sotnikov} where excitons correspond to the LS to high-spin (HS) excitations.\cite{KunesJPCM, Nasu}
To elucidate this possibility, it is plausible to seek for further magnetic transitions up to 1000 T, up to which several exotic phase transitions and a re-entrant to another excitonic condensation are predicted.\cite{Tatsuno} Such experiment should be possible using the current technique in combination with the electro-magnetic flux compression technique.\cite{TakeyamaEMFC}
Such measurements are now underway.}

\section{Conclusion}

A high-speed 100 MHz strain monitor  was devised using an FBG-based strain monitor and optical filter, whose resolution was approximately $\Delta L/L\sim10^{-4}$.
With the devised strain monitor, the longitudinal magnetostriction of LaCoO$_{3}$ was measured at several cryogenic temperatures up to a maximum magnetic field of 150 T.
The system is simple to operate and more compact than the conventional ones that have been reported. \cite{Rodriguez}
\textbl{Magnetostriction measurements at $B$>100 T has been accomplished for the first time.
The system is readily compatible with even higher magnetic field up to 700 T generated with the electromagnetic flux compression technique.\cite{TakeyamaEMFC}}
Further improvement in the signal-to-noise ratio is plausible for the magnetostriction measurements of quantum magnets.

\begin{acknowledgments}
This work was supported by JSPS KAKENHI Grant-in-Aid for Young Scientists (B) Grant No. 16K17738, Grant-in-Aid for Scientific Research (B) Grant No. 16H04009 and the internal research grant from ISSP, UTokyo.
The X-ray facility in ISSP, UTokyo is acknowledged for its help in determining the crystalline axes.
Koshin Kogaku Co., Ltd. is acknowledged for the technical advises on the optical filters. 
S. Minakuchi (Univ. of Tokyo, Japan) and H. Tsuda (AIST, Japan) are acknowledged for the technical advices on FBG. 
\end{acknowledgments}


\providecommand{\noopsort}[1]{}\providecommand{\singleletter}[1]{#1}%

\end{document}